\documentclass{article}
\usepackage{spconf,amsmath}
\usepackage[pdftex]{graphicx}
\usepackage{color}
\usepackage{url}
\usepackage{hyperref}
\usepackage{caption}
\usepackage{balance}

\renewenvironment{thebibliography}[1]{%
\begin{oldthebibliography}{#1}%
\setlength{\itemsep}{-.3ex}%
}%
{%
\end{oldthebibliography}%
}

\title{CNN Architectures for Large-Scale Audio Classification}
\makeatletter
\def\@name{ \emph{Shawn Hershey, Sourish Chaudhuri, Daniel P. W. Ellis, Jort F. Gemmeke, Aren Jansen, R. Channing Moore, }  \\ \emph{Manoj Plakal, Devin Platt, Rif A. Saurous, Bryan Seybold, Malcolm Slaney, Ron J. Weiss, Kevin Wilson}\vspace{6pt}}
\makeatother

\address{Google, Inc., New York, NY, and Mountain View, CA, USA\\
{\footnotesize \tt shershey@google.com}}

\begin{document}
\ninept

\maketitle

\begin{abstract}
Convolutional Neural Networks (CNNs) have proven very effective in image
classification and show promise for audio.  We use various CNN architectures
to classify the soundtracks of a dataset of 70M
training videos (5.24 million hours) with 30,871 video-level labels.  We examine
fully connected Deep Neural Networks (DNNs), AlexNet~\cite{krizhevsky2012imagenet},
VGG~\cite{simonyan2014very},
Inception~\cite{szegedy2015rethinking}, and ResNet~\cite{he2015deep}.
We investigate varying the size of both training set and label vocabulary,
finding that analogs of the CNNs used in image classification do well
on our audio classification task, and larger training and label sets
help up to a point.  A model using embeddings from these classifiers
does much better than raw features on the {\em Audio Set}~\cite{audioset}
Acoustic Event Detection (AED) classification task.
\end{abstract}
\section{Introduction}
\label{sec:intro}
Image classification performance has improved greatly with the advent of large datasets
such as ImageNet~\cite{deng2009imagenet} using Convolutional Neural Network (CNN)
architectures such as
AlexNet~\cite{krizhevsky2012imagenet}, VGG~\cite{simonyan2014very},
Inception~\cite{szegedy2015rethinking}, and ResNet~\cite{he2015deep}.  We are
curious to see if similarly large datasets and CNNs can yield good performance on audio
classification problems.  Our dataset consists of 70 million (henceforth 70M)
training videos totalling 5.24 million hours, each tagged from a set of 30,871 (henceforth 30K)
labels.  We call this dataset YouTube-100M.  Our primary task is to predict the
video-level labels using audio information (i.e., soundtrack classification).
Per Lyon~\cite{lyon2010machine},
teaching machines to hear and understand video can improve our ability to
``categorize, organize, and index them''.

In this paper, we use the YouTube-100M dataset to investigate:
how popular Deep Neural Network (DNN) architectures compare on video soundtrack classification;
how performance varies with different training set and label vocabulary sizes; and
whether our trained models can also be useful for AED.

Historically, AED has been addressed with features such as
MFCCs and classifiers based on GMMs, HMMs, NMF, or SVMs ~\cite{mesaros2010acoustic,zhuang2010real, gemmeke2013exemplar, temko2006clear}.
More recent approaches use some form of DNN, including CNNs~\cite{takahashi2016deep}
and RNNs~\cite{parascandolo2016recurrent}.
Prior work has been reported on datasets such as TRECVid~\cite{2016trecvidawad},
ActivityNet~\cite{caba2015activitynet}, Sports1M~\cite{KarpathyCVPR14},
and TUT/DCASE Acoustic scenes 2016~\cite{DCASE2016}
which are much smaller than YouTube-100M.  Our large dataset puts us in
a good position to evaluate models with large model capacity.

RNNs and CNNs have been used in Large Vocabulary Continuous Speech
Recognition (LVCSR)~\cite{sainath2015convolutional}.  Unlike that task,
our labels apply to entire videos without any changes in time, so we
have yet to try such recurrent models.

Eghbal-Zadeh et al.~\cite{eghbalcp} recently won the DCASE 2016
Acoustic Scene Classification (ASC) task, which, like soundtrack classification,
involves assigning a single label to an audio clip containing many
events.  Their system used spectrogram features feeding a VGG classifier,
similar to one of the classifiers in our work.  This paper, however,
compares the performance of several different architectures.  To our knowledge,
we are the first to publish results of Inception and ResNet networks applied to
audio.

We aggregate local classifications to whole-soundtrack decisions by imitating
the visual-based video classification of Ng et al.~\cite{yue2015beyond}.  After
investigating several more complex models for combining information across time,
they found simple averaging of single-frame CNN classification outputs performed
nearly as well.  By analogy, we apply a classifier to a series of
non-overlapping segments, then average all the sets of classifier outputs.

Kumar et al.~\cite{kumar2016audio} consider AED in a dataset with
video-level labels as a Multiple Instance Learning (MIL)
problem, but remark that scaling such approaches remains an open problem.  By
contrast, we are investigating the limits of training with weak labels for
very large datasets. While many of the individual segments will be uninformative
about the labels inherited from the parent video, we hope that,
given enough training, the net can learn to spot useful cues.
We are not able to quantify how ``weak'' the labels are (i.e., what
proportion of the segments are uninformative), and for the majority of classes
(e.g., ``Computer Hardware'', ``Boeing 757'', ``Ollie''), it's not clear how
to decide relevance.  Note that for some classes (e.g. ``Beach''),
background ambiance is itself informative.

Our dataset size allows us to examine networks
with large model capacity, fully exploiting ideas from the image classification literature.
By computing log-mel spectrograms of multiple frames, we
create 2D image-like patches to present to the classifiers.
Although the distinct meanings of time and frequency axes might argue
for audio-specific architectures, this work employs
minimally-altered image classification networks such as Inception-V3 and
ResNet-50.  We train with subsets of YouTube-100M spanning 23K to 70M videos
to evaluate
the impact of training set size on performance, and we investigate the effects
of label set size on generalization by training
models with subsets of labels, spanning 400 to 30K, which are then evaluated on a single
common subset of labels.  We additionally examine the usefulness of our
networks for AED by examining the performance of a model trained with
embeddings from one of our networks on the {\em Audio Set}~\cite{audioset} dataset.

\section{Dataset}
\label{sec:dataset}

The YouTube-100M data set consists of 100 million YouTube videos: 70M training
videos, 10M
evaluation videos, and a pool of 20M videos that we use
for validation.  Videos average 4.6 minutes each for a total of 5.4M training hours.  Each of these videos is
labeled with 1 or more topic identifiers (from Knowledge Graph \cite{singhal2012introducing})
from a set of 30,871 labels.  There are an average of around 5 labels
per video.  The labels are
assigned automatically based on a combination of metadata (title, description,
comments, etc.), context, and image content for each video.
The labels apply to the entire video and range
from very generic (e.g. ``Song'') to very specific (e.g. ``Cormorant'').  Table ~\ref{table:example_labels}
shows a few examples.

\begin{table}[t]
  \caption{Example labels from the 30K set.}
  \label{table:example_labels}
  \begin{tabular}{r p{6cm}}
    \hline
    Label prior & Example Labels \\ \hline
    $0.1 \ldots 0.2$ & Song, Music, Game, Sports, Performance \\
    $0.01 \ldots 0.1$ & Singing, Car, Chordophone, Speech \\
    $\sim 10^{-5}$ & Custom Motorcycle, Retaining Wall \\
    $\sim 10^{-6}$ & Cormorant, Lecturer \\
    \hline
  \end{tabular}
\end{table}

Being machine generated, the labels are not 100\%  accurate
and of the 30K labels, some are clearly acoustically relevant (``Trumpet'')
and others are less so (``Web Page'').  Videos often bear annotations with multiple
degrees of specificity.  For example, videos labeled with ``Trumpet''
are often labeled ``Entertainment'' as well, although no hierarchy is enforced.

\section{Experimental Framework}
\label{sec:framework}

\subsection{Training}
\label{sec:training}
The audio is divided into non-overlapping 960~ms frames.
This gave approximately 20 billion examples from the 70M videos.
Each frame inherits all the labels of its parent video.
The 960~ms frames are
decomposed with a short-time Fourier transform applying 25~ms windows every 10~ms.
The resulting spectrogram is integrated into 64 mel-spaced frequency bins, and
the magnitude of each bin is log-transformed after adding a small offset to avoid
numerical issues.  This gives log-mel spectrogram patches of $96 \times 64$ bins
that form the input to all classifiers.  During training we fetch mini-batches
of 128 input examples by randomly sampling from all patches.

All experiments used TensorFlow~\cite{tensorflow2015-whitepaper} and were trained
asynchronously on multiple GPUs using the Adam~\cite{kingma2014adam}
optimizer.  We performed grid searches over learning rates, batch sizes, number
of GPUs, and parameter servers.  Batch normalization~\cite{ioffe2015batch}
was applied after all convolutional layers.  All models used a final sigmoid layer rather than a softmax layer since
each example can have multiple labels.  Cross-entropy was the loss function.
In view of the large training set size, we did not use dropout~\cite{srivastava2014dropout},
weight decay,
or other common regularization techniques.  For the models trained on 7M
or more examples, we saw no evidence of overfitting.  During training, we
monitored progress via 1-best accuracy and mean Average Precision (mAP)
over a validation subset.

\subsection{Evaluation}
\label{sec:Evaluation}
From the pool of 10M evaluation videos we created three balanced
evaluation sets, each with roughly 33 examples per class: 1M videos
for the 30K labels, 100K videos for the 3087 (henceforth 3K) most frequent
labels, and 12K
videos for the 400 most frequent labels.
%
%To calculate video-level probabilities for each class,
We passed each 960~ms frame from each evaluation video
through the classifier.  We then averaged the classifier output scores % for each class
across all segments in a video.

For our metrics, we calculated the balanced average across all classes of AUC (also reported as the equivalent d-prime
class separation), and mean Average Precision (mAP). AUC is the area under the
Receiver Operating Characteristic (ROC) curve
\cite{fawcett2004roc}, that is, the probability of correctly classifying a
positive example (correct accept rate) as a function of the probability of
incorrectly classifying a negative example as positive (false accept rate);
perfect classification achieves AUC of 1.0 (corresponding to an infinite d-prime),
and random guessing gives an AUC of 0.5 (d-prime of zero).\footnote{$d' = \sqrt{2} F^{-1}(\text{AUC})$ where $F^{-1}$ is
the inverse cumulative distribution function for a unit Gaussian.}  mAP is the
mean across classes of the Average Precision (AP), which is the proportion of
positive items in a ranked list of trials (i.e., Precision) averaged across
lists just long enough to include each individual positive trial
\cite{buckley2004retrieval}.  AP is widely used as an indicator of
precision that does not require a particular retrieval list length, but,
unlike AUC, it is directly correlated with the prior probability of the class.
Because most of our classes have very low priors
($<10^{-4}$), the mAPs we report are typically small, even
though the false alarm rates are good.

\subsection{Architectures}
\label{sec:framework_architectures}

Our baseline is a fully connected DNN, which we compared to several
networks closely modeled on successful image classifiers.
For our baseline experiments,
we trained and evaluated using only the 10\% most
frequent labels of the original 30K (i.e, 3K labels). For each experiment, we
coarsely optimized number of GPUs and learning rate for the
frame level classification accuracy.  The optimal
number of GPUs represents a compromise between overall computing power and
communication overhead, and varies by architecture.

\subsubsection{Fully Connected}
\label{sec:fullyconnected}
Our baseline network is a fully connected model with RELU activations~\cite{nair2010rectified},
$N$ layers, and $M$ units per
layer.  We swept over $N = [2, 3, 4, 5, 6]$ and $M = [500, 1000, 2000, 3000, 4000]$.
Our best performing model had $N = 3$
layers, $M = 1000$ units, learning rate of $3 \times 10^{-5}$, 10 GPUs and 5 parameter servers.
This network has approximately 11.2M weights and 11.2M multiplies.

\subsubsection{AlexNet}
\label{sec:alexnet}
The original AlexNet~\cite{krizhevsky2012imagenet} architectures was designed for a $224 \times 224 \times 3$ input with an
initial $11 \times 11$ convolutional layer with a stride of 4.  Because our inputs
are $96 \times 64$, we use a stride of $2 \times 1$ so that the number of activations
are similar after the initial layer.   We also use batch normalization after each
convolutional layer instead of local response normalization (LRN) and
replace the final 1000 unit layer with a 3087 unit layer.  While the original
AlexNet has approximately 62.4M weights and 1.1G
multiplies, our version has 37.3M weights and 767M multiplies.  Also, for
simplicity, unlike the original AlexNet, we do not split filters across multiple
devices.  We trained with 20 GPUs and 10 parameter servers.

\subsubsection{VGG}
\label{sec:vgg}
The only changes we made to VGG (configuration E)~\cite{simonyan2014very} were to the final layer (3087 units with a
sigmoid) as well as the use of batch normalization instead of LRN.  While the
original network had 144M weights and 20B multiplies,
the audio variant uses 62M weights and 2.4B multiplies.  We tried another variant
that reduced the initial strides (as we did with AlexNet), but found that not
modifying the strides resulted in faster training and better performance.
With our setup, parallelizing beyond 10 GPUs did not help significantly, so we
trained with 10 GPUs and 5 parameter servers.

\subsubsection{Inception V3}
\label{sec:inception}
We modified the inception V3~\cite{szegedy2015rethinking} network by removing the first four layers of the
stem, up to and including the MaxPool, as well as removing the auxiliary network.
We changed the Average Pool size to $10 \times 6$ to reflect the change in activations.
We tried including the stem and removing the first stride of 2 and MaxPool but
found that it performed worse than the variant with the truncated stem.
The original network has 27M weights with 5.6B multiplies, and the audio variant
has 28M weights and 4.7B multiplies.  We trained with 40 GPUs and 20 parameter servers.

\subsubsection{ResNet-50}
\label{sec:resnet}
We modified ResNet-50~\cite{he2015deep} by removing the stride of 2 from the first $7 \times 7$ convolution so
that the number of activations was not too different in the audio version.
We changed the Average Pool size to $6 \times 4$ to reflect the change in activations.
The original network has 26M weights and 3.8B multiplies.
The audio variant has 30M
weights and 1.9B multiplies.  We trained with 20 GPUs
and 10 parameter servers.

\section{Experiments}
\label{sec:experiments}

\begin{table}[t]
  \caption{Comparison of performance of several DNN architectures trained on
  70M videos, each tagged with labels from a set of 3K.  The last row contains
  results for a model that was trained much longer than the others, with a
  reduction in learning rate after 13 million steps.}
  \label{table:arch_val_metrics}
  \begin{tabular}{ l l l l l p{1cm}}
    \hline
    Architectures & Steps & Time & AUC & d-prime & mAP \\ \hline
    Fully Connected & 5M & 35h & 0.851 & 1.471 & 0.058 \\
    AlexNet & 5M & 82h & 0.894 & 1.764 & 0.115 \\
    VGG & 5M & 184h & 0.911 & 1.909 & 0.161 \\
    Inception V3 & 5M & 137h & \textbf{0.918} & \textbf{1.969} & 0.181 \\
    ResNet-50 & 5M & 119h & 0.916 & 1.952 & \textbf{0.182}  \\ \hline
    ResNet-50 & 17M & 356h & \textbf{0.926} & \textbf{2.041} & \textbf{0.212}  \\
    \hline
  \end{tabular}
\end{table}

\begin{table*}[t!]
  \begin{minipage}{.5\linewidth}
    \captionsetup{width=0.9\textwidth}
    \caption{Results of varying label set size, evaluated over 400 labels.
    All models are variants of ResNet-50 trained on 70M videos.  The bottleneck,
    if present, is 128 dimensions.}
    \label{table:eval_400_labels}
    \centering
    \begin{tabular}{l l l l l}
      \hline
      Bneck & Labels & AUC & d-prime & mAP \\ \hline
      no & 30K & --- & --- & --- \\
      no & 3K & \textbf{0.930} & \textbf{2.087} & \textbf{0.381} \\
      no & 400 & 0.928 & 2.067 & 0.376 \\
      yes & 30K & 0.925 & 2.035 & 0.369 \\
      yes & 3K & 0.919 & 1.982 & 0.347 \\
      yes & 400 & 0.924 & 2.026 & 0.365 \\
      \hline
    \end{tabular}
  \end{minipage}%
  \begin{minipage}{0.5\linewidth}
    \captionsetup{width=0.9\textwidth}
    \caption{Results of training with different amounts of data.  All rows used
    the same ResNet-50 architecture trained on videos tagged with labels from a set of 3K.}
    \label{table:sweep_training}
    \centering
    \begin{tabular}{l l l l}
      \hline
      Training Videos & AUC & d-prime & mAP \\ \hline
      70M & \textbf{0.923} & \textbf{2.019} & \textbf{0.206} \\
      7M & 0.922 & 2.006 & 0.202 \\
      700K & 0.921 & 1.997 & 0.203 \\
      70K & 0.909 & 1.883 & 0.162 \\
      23K & 0.868 & 1.581 & 0.118 \\
      \hline
    \end{tabular}
  \end{minipage}
\end{table*}

\begin{figure}[t!]
\centering
\includegraphics[width=8.5cm]{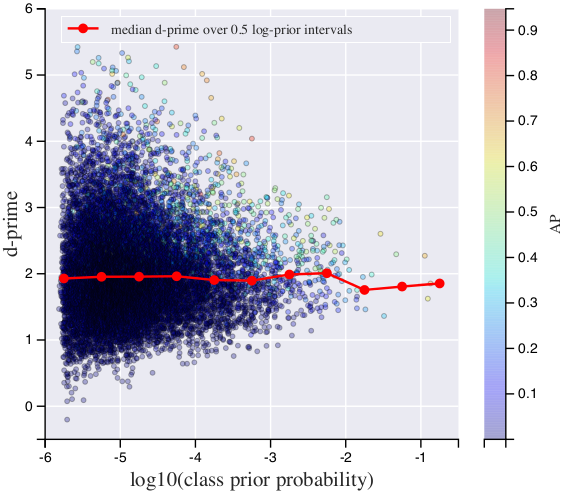}
\caption{Scatter plot of ResNet-50's per-class d-prime versus log prior probability.  Each
point is a separate class from a random 20\% subset of the 30K set.  Color reflects the class AP.}
\label{fig:prior_dprime_scatter}
\end{figure}

\begin{figure*}[t!]
%\begin{figure*}[ht!]
  \centering
  \begin{tabular}{ccc}
    Trumpet & Piano & Guitar \\
    \includegraphics[width=2.2in]{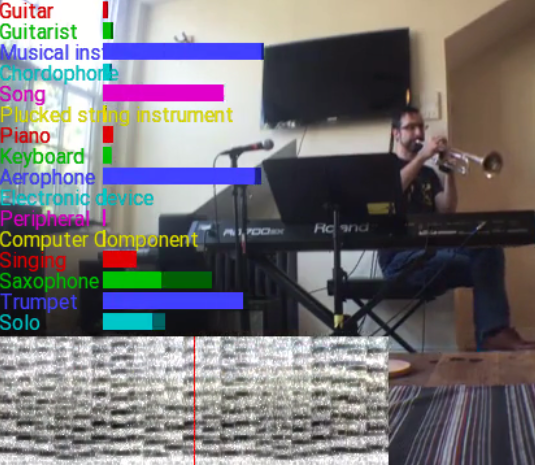} &
    \includegraphics[width=2.2in]{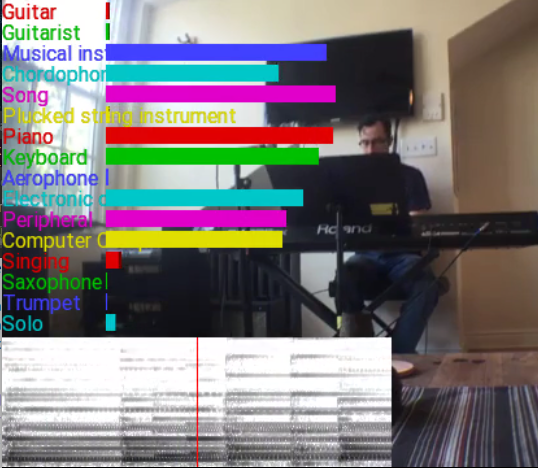} &
    \includegraphics[width=2.2in]{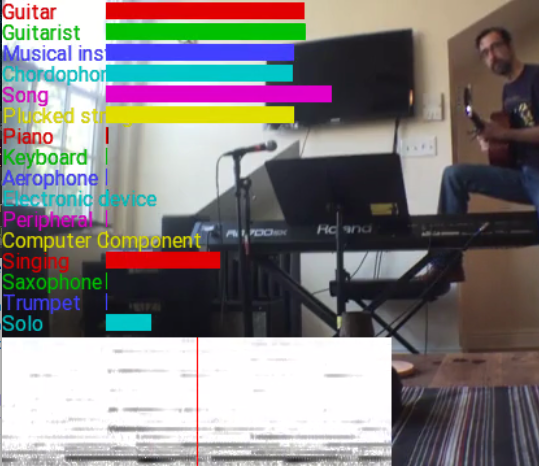} \\
  \end{tabular}
  \caption{Three example excerpts from a video classified by ResNet-50 with instantaneous model outputs overlaid.
  The 16 classifier outputs with the greatest peak values across the entire video were chosen from the 30K set for display.
  }
  \label{fig:examples}
\end{figure*}

\subsection{Architecture Comparison}
\label{sec:architecture_comparison}

For all network architectures we trained with 3K labels and 70M videos and compared after 5 million
mini-batches of 128 inputs.
Because some networks trained faster than others, comparing after a fixed
wall-clock time would give slightly different results but would not change
the relative ordering of the architectures' performance.  We include numbers
for ResNet after training for 17 million mini-batches (405 hours) to show that
performance continues to improve.  We reduced the learning rate by a factor
of 10 after 13 million mini-batches.

Table ~\ref{table:arch_val_metrics} shows the evaluation results calculated
over the 100K balanced videos.   All CNNs beat the fully-connected baseline.
Inception and ResNet achieve the best performance;
they provide high model capacity and their convolutional units can efficiently
capture common structures that may occur in different areas of the input array
for both images, and, we infer, our audio representation.

To investigate how the prior likelihood of each label affects its
performance,  Fig.~\ref{fig:prior_dprime_scatter} shows a scatter
plot of the 30K classes with label frequency on the $x$ axis and ResNet-50's d-prime on the $y$ axis.
d-prime seems to stay centered around 2.0 across label prior,
although the variance of d-prime increases for less-common classes. This is
contrary to the usual result where classifier performance improves with
increased training data, particularly over the 5 orders of magnitude
illustrated in the plot.

\subsection{Label Set Size}
\label{sec:labelsetsize}

Using a 400 label subset of the 30K labels, we investigated how training with
different subsets of classes can affect
performance, perhaps by encouraging intermediate representations that better generalize
to unseen examples even for the evaluation classes.  In addition
to examining three label set sizes (30K, 3K, and 400), we compared models with and
without a bottleneck layer of 128 units placed right before the final output
layer.  We introduced the bottleneck layer to speed up the training of the model
trained with 30K labels.  Without a bottleneck, the larger output layer
increased the number of weights from 30M to 80M and significantly reduced training
speed.  We do not report metrics on the 30K label
model without the bottleneck because it would have taken several months to train.
For all label set size experiments, we used
the ResNet-50 model and trained for 5 million mini-batches of 128 inputs (about 120 hours) on 70M videos.

Tables ~\ref{table:eval_400_labels} shows the results.
When comparing models with the bottleneck, we see that performance does indeed
improve slightly as we increase the number of labels we trained on, although
networks without the bottleneck have higher performance overall.  The bottleneck
layer is relatively small compared to the 2048 activations coming out of ResNet-50's
Average Pool layer and so it is effecting a substantial reduction in information.
These results provide weak support to the notion that training with a broader
set of categories can help to regularize even the 400 class subset.

\subsection{Training Set Size}
\label{sec:training_set_size}
Having a very large training set available allows us to investigate
how training set size affects performance.
With 70M videos and an average of 4.6 minutes
per video, we have around 20 billion 960~ms training examples.  Given ResNet-50's
training speed of 11 mini-batches per second with 20 GPUs, it would take 23 weeks for the
network to see each pattern once (one epoch).
However, if all videos were equal length and fully randomized, we expect to see
at least one frame from each video in only 14 hours.
We hypothesize that, even if we cannot get through an entire epoch, 70M
videos will provide an advantage over 7M by virtue of the greater diversity
of videos underlying the limited number of training patterns consumed.
We trained a ResNet-50 model for 16 million mini-batches of 128 inputs  (about 380 hours) on the 3K label
set with 70M, 7M, 700K, 70K, and 23K videos.

The video level results are shown in Table ~\ref{table:sweep_training}.
The 70K and 23K models show worse performance but the validation plots (not included)
showed that they likely suffered from
overfitting.  Regularization techniques (or data augmentation) might have
boosted the numbers on these smaller training sets.
The 700K, 7M, and 70M models are mostly very close in performance
although the 700K model is slightly inferior.

\subsection{AED with the Audio Set Dataset}
\label{sec:audio_set_aed}

{\em Audio Set}~\cite{audioset} is a dataset of over 1 million 10~second excerpts
labeled with a vocabulary of acoustic events (whereas not all of the
YouTube-100M 30K labels pertain to acoustic events).  This comes to about 3000 hours --
still only $\approx 0.05\%$ of YouTube-100M.
We train two fully-connected models to
predict labels for {\em Audio Set}.  The first model uses $64 \times 20$ log-mel patches
and the second uses the output of the penultimate ``embedding'' layer of our best ResNet
model as inputs.   The log-mel baseline achieves a balanced mAP of 0.137 and AUC
of 0.904 (equivalent to d-prime of 1.846). The model trained on embeddings
achieves mAP / AUC / d-prime of 0.314 / 0.959 / 2.452. This jump in performance
reflects the benefit of the larger YouTube-100M training set embodied in the
ResNet classifier outputs.

\balance

\section{Conclusions}

The results in Section ~\ref{sec:architecture_comparison} show that
state-of-the-art image networks are capable of excellent results on
audio classification when compared to a simple fully connected network
or earlier image classification architectures.  In Section ~\ref{sec:labelsetsize} we saw results showing that training on
larger label set vocabularies can improve performance, albeit modestly, when
evaluating on smaller label sets.  In Section ~\ref{sec:training_set_size} we saw that increasing the number of videos up
to 7M improves performance for the best-performing ResNet-50 architecture.
We note that regularization could have reduced the
gap between the models trained on smaller datasets and the 7M and 70M datasets.
In Section ~\ref{sec:audio_set_aed} we see a significant increase over our baseline
when training a model for AED with ResNet embeddings on the {\em Audio Set} dataset.

In addition to these quantified results, we can subjectively examine the
performance of the model on segments of video.  Fig.~\ref{fig:examples}
shows the results of running our best classifier over a video and
overlaying the frame-by-frame results of the 16 classifier outputs with the greatest
peak values across the entire video.  The different sound sources present
at different points in the video are clearly distinguished.
\footnote{A similar video is available online at \href{https://youtu.be/oAAo_r7ZT8U}{https://youtu.be/oAAo\_r7ZT8U}.}

\section{Acknowledgements}
The authors would like to thank George Toderici and Marvin Ritter, both with Google, for their very valuable feedback.

\vfill\pagebreak

\bibliographystyle{IEEEbib}
\bibliography{refs}\label{sec:refs}

\begin{thebibliography}{10}

\bibitem{krizhevsky2012imagenet}
A.~Krizhevsky, I.~Sutskever, and G.~E. Hinton,
\newblock ``Imagenet classification with deep convolutional neural networks,''
\newblock in {\em Advances in neural information processing systems}, 2012, pp.
  1097--1105.

\bibitem{simonyan2014very}
K.~Simonyan and A.~Zisserman,
\newblock ``Very deep convolutional networks for large-scale image
  recognition,''
\newblock {\em arXiv preprint arXiv:1409.1556}, 2014.

\bibitem{szegedy2015rethinking}
C.~Szegedy, V.~Vanhoucke, S.~Ioffe, J.~Shlens, and Z.~Wojna,
\newblock ``Rethinking the inception architecture for computer vision,''
\newblock {\em arXiv preprint arXiv:1512.00567}, 2015.

\bibitem{he2015deep}
K.~He, X.~Zhang, S.~Ren, and J.~Sun,
\newblock ``Deep residual learning for image recognition,''
\newblock {\em arXiv preprint arXiv:1512.03385}, 2015.

\bibitem{audioset}
J.~F. Gemmeke, D.~P.~W. Ellis, D.~Freedman, A.~Jansen, W.~Lawrence, R.~C.
  Moore, M.~Plakal, and M.~Ritter,
\newblock ``Audio {S}et: An ontology and human-labeled dartaset for audio
  events,''
\newblock in {\em IEEE ICASSP 2017}, New Orleans, 2017.

\bibitem{deng2009imagenet}
J.~Deng, W.~Dong, R.~Socher, L.-J. Li, K.~Li, and L.~Fei-Fei,
\newblock ``Imagenet: A large-scale hierarchical image database,''
\newblock in {\em Computer Vision and Pattern Recognition, 2009. CVPR 2009.
  IEEE Conference on}. IEEE, 2009, pp. 248--255.

\bibitem{lyon2010machine}
R.~F. Lyon,
\newblock ``Machine hearing: An emerging field [exploratory dsp],''
\newblock {\em Ieee signal processing magazine}, vol. 27, no. 5, pp. 131--139,
  2010.

\bibitem{mesaros2010acoustic}
A.~Mesaros, T.~Heittola, A.~Eronen, and T.~Virtanen,
\newblock ``Acoustic event detection in real life recordings,''
\newblock in {\em Signal Processing Conference, 2010 18th European}. IEEE,
  2010, pp. 1267--1271.

\bibitem{zhuang2010real}
X.~Zhuang, X.~Zhou, M.~A. Hasegawa-Johnson, and T.~S. Huang,
\newblock ``Real-world acoustic event detection,''
\newblock {\em Pattern Recognition Letters}, vol. 31, no. 12, pp. 1543--1551,
  2010.

\bibitem{gemmeke2013exemplar}
J.~F. Gemmeke, L.~Vuegen, P.~Karsmakers, B.~Vanrumste, et~al.,
\newblock ``An exemplar-based nmf approach to audio event detection,''
\newblock in {\em 2013 IEEE Workshop on Applications of Signal Processing to
  Audio and Acoustics}. IEEE, 2013, pp. 1--4.

\bibitem{temko2006clear}
A.~Temko, R.~Malkin, C.~Zieger, D.~Macho, C.~Nadeu, and M.~Omologo,
\newblock ``Clear evaluation of acoustic event detection and classification
  systems,''
\newblock in {\em International Evaluation Workshop on Classification of
  Events, Activities and Relationships}. Springer, 2006, pp. 311--322.

\bibitem{takahashi2016deep}
N.~Takahashi, M.~Gygli, B.~Pfister, and L.~Van~Gool,
\newblock ``Deep convolutional neural networks and data augmentation for
  acoustic event detection,''
\newblock {\em arXiv preprint arXiv:1604.07160}, 2016.

\bibitem{parascandolo2016recurrent}
G.~Parascandolo, H.~Huttunen, and T.~Virtanen,
\newblock ``Recurrent neural networks for polyphonic sound event detection in
  real life recordings,''
\newblock in {\em 2016 IEEE International Conference on Acoustics, Speech and
  Signal Processing (ICASSP)}. IEEE, 2016, pp. 6440--6444.

\bibitem{2016trecvidawad}
G.~Awad, J.~Fiscus, M.~Michel, D.~Joy, W.~Kraaij, A.~F. Smeaton, G.~Quéenot,
  M.~Eskevich, R.~Aly, and R.~Ordelman,
\newblock ``Trecvid 2016: Evaluating video search, video event detection,
  localization, and hyperlinking,''
\newblock in {\em Proceedings of TRECVID 2016}. NIST, USA, 2016.

\bibitem{caba2015activitynet}
B.~G. Fabian Caba~Heilbron, Victor~Escorcia and J.~C. Niebles,
\newblock ``Activitynet: A large-scale video benchmark for human activity
  understanding,''
\newblock in {\em Proceedings of the IEEE Conference on Computer Vision and
  Pattern Recognition}, 2015, pp. 961--970.

\bibitem{KarpathyCVPR14}
A.~Karpathy, G.~Toderici, S.~Shetty, T.~Leung, R.~Sukthankar, and L.~Fei-Fei,
\newblock ``Large-scale video classification with convolutional neural
  networks,''
\newblock in {\em CVPR}, 2014.

\bibitem{DCASE2016}
A.~Mesaros, T.~Heittola, and T.~Virtanen,
\newblock ``{TUT} database for acoustic scene classification and sound event
  detection,''
\newblock in {\em 24th European Signal Processing Conference 2016 (EUSIPCO
  2016)}, Budapest, Hungary, 2016,
\newblock \url{http://www.cs.tut.fi/sgn/arg/dcase2016/}.

\bibitem{sainath2015convolutional}
T.~N. Sainath, O.~Vinyals, A.~Senior, and H.~Sak,
\newblock ``Convolutional, long short-term memory, fully connected deep neural
  networks,''
\newblock in {\em 2015 IEEE International Conference on Acoustics, Speech and
  Signal Processing (ICASSP)}. IEEE, 2015, pp. 4580--4584.

\bibitem{eghbalcp}
H.~Eghbal-Zadeh, B.~Lehner, M.~Dorfer, and G.~Widmer,
\newblock ``Cp-jku submissions for dcase-2016: A hybrid approach using binaural
  i-vectors and deep convolutional neural networks,''
\newblock .

\bibitem{yue2015beyond}
J.~Yue-Hei~Ng, M.~Hausknecht, S.~Vijayanarasimhan, O.~Vinyals, R.~Monga, and
  G.~Toderici,
\newblock ``Beyond short snippets: Deep networks for video classification,''
\newblock in {\em Proceedings of the IEEE Conference on Computer Vision and
  Pattern Recognition}, 2015, pp. 4694--4702.

\bibitem{kumar2016audio}
A.~Kumar and B.~Raj,
\newblock ``Audio event detection using weakly labeled data,''
\newblock {\em arXiv preprint arXiv:1605.02401}, 2016.

\bibitem{singhal2012introducing}
A.~Singhal,
\newblock ``Introducing the knowledge graph: things, not strings,'' 2012,
\newblock Official Google blog,
  \url{https://googleblog.blogspot.com/2012/05/introducing-knowledge-graph-things-not.html}.

\bibitem{tensorflow2015-whitepaper}
M.~Abadi et~al.,
\newblock ``{TensorFlow}: Large-scale machine learning on heterogeneous
  systems,'' 2015,
\newblock Software available from tensorflow.org.

\bibitem{kingma2014adam}
D.~Kingma and J.~Ba,
\newblock ``Adam: A method for stochastic optimization,''
\newblock {\em arXiv preprint arXiv:1412.6980}, 2014.

\bibitem{ioffe2015batch}
S.~Ioffe and C.~Szegedy,
\newblock ``Batch normalization: Accelerating deep network training by reducing
  internal covariate shift,''
\newblock {\em arXiv preprint arXiv:1502.03167}, 2015.

\bibitem{srivastava2014dropout}
N.~Srivastava, G.~E. Hinton, A.~Krizhevsky, I.~Sutskever, and R.~Salakhutdinov,
\newblock ``Dropout: a simple way to prevent neural networks from
  overfitting.,''
\newblock {\em Journal of Machine Learning Research}, vol. 15, no. 1, pp.
  1929--1958, 2014.

\bibitem{fawcett2004roc}
T.~Fawcett,
\newblock ``Roc graphs: Notes and practical considerations for researchers,''
\newblock {\em Machine learning}, vol. 31, no. 1, pp. 1--38, 2004.

\bibitem{buckley2004retrieval}
C.~Buckley and E.~M. Voorhees,
\newblock ``Retrieval evaluation with incomplete information,''
\newblock in {\em Proceedings of the 27th annual international ACM SIGIR
  conference on Research and development in information retrieval}. ACM, 2004,
  pp. 25--32.

\bibitem{nair2010rectified}
V.~Nair and G.~E. Hinton,
\newblock ``Rectified linear units improve restricted boltzmann machines,''
\newblock in {\em Proceedings of the 27th International Conference on Machine
  Learning (ICML-10)}, 2010, pp. 807--814.

\end{thebibliography}

\end{document}